\documentclass{mem}
\usepackage{natbib}
\usepackage{txfonts}
\usepackage{balance}
\usepackage{graphicx}
\idline{75}{282}
\begin{document}
\def\teff{$T\rm_{eff }$}
\def\kms{$\mathrm {km s}^{-1}$}

\title{
A multifrequency view of starburst galaxies
}

   \subtitle{}

\author{
J.\ K.\ Becker\inst{1} 
\and
F.\ Schuppan\inst{1}
\and S.\ Sch\"oneberg\inst{1}
          }

  \offprints{J.\ K.\ Becker}

\institute{
Institut f\"ur Theoretische Physik, Fakult\"at f\"ur Physik \& Astronomie,
Ruhr-Universit\"at Bochum, 44780 Bochum
\email{julia.becker@rub.de}
}

\authorrunning{Becker}

\titlerunning{Starburst galaxies}

\abstract{
During the past few years, first observations of starburst galaxies at
$>$~GeV energies could be made with the Fermi Gamma-ray Space
Telescope (GeV range) and Imaging Air Cherenkov Telescopes (TeV
range). The two nearest starbursts, M82 and NGC253 were detected, and
most recently, the detection of two starburst-Seyfert composites (NGC1068 and
NGC4945) were reported. The emission for the two starbursts is best explained by hadronic interactions, and
thus providing a first, unique opportunity to study the role of cosmic rays
in galaxies. In this paper, the role of cosmic rays for the
non-thermal component of galaxies is reviewed by discussing the entire
non-thermal frequency range from radio emission to TeV energies. In particular, the interpretation of radio emission 
arising from electron synchrotron radiation is predicted to be correlated to TeV emission coming from interactions of accelerated
hadrons. This is 
observed for the few objects known at TeV energies, but the correlation needs to be established with
significantly higher statistics. An
outlook of the possibility of tracing cosmic rays with molecular ions
is given.
\keywords{starburst galaxies --
high-energy photons --cosmic rays -- molecular ions}
}
\maketitle{}

\section{Introduction}

Galaxies with a high star formation rate provide the opportunity to
study the influence of massive stars on the large-scale behavior of
galaxies in detail. In particular, the role of gas heating by star
formation can be studied by the observation of far-infrared emission
and the role of supernova remnants can be investigated by looking at
non-thermal radio emission from electron synchrotron radiation. Most
recently, first gamma-ray detections using the Fermi Gamma-ray Space
Telescope (FGST) and Imaging Air Cherenkov Telescopes like
H.E.S.S.\ and VERITAS were made. The nearest starburst galaxies M82
and NGC253 were detected above GeV energies, tracing the interaction
of hadronic cosmic rays with the gas in the galaxy. 

In this paper, the multifrequency spectrum of starburst galaxies will
be reviewed. In Section \ref{radio_fir}, the correlation between the
far infrared and radio emission is discussed. In Section
\ref{proton_proton:sec}, hadronic interactions at energies $E>$~GeV
are discussed in the context of neutral secondary production. 
In Section 
\ref{crtracers:sec}, cosmic ray-induced ionization is mentioned as a
future method of tracing hadronic cosmic rays in galactic
environments. In particular,  a supernova remnant interacting with a molecular cloud
provides an optimal environment for the production of H$_{2}^{+}$, at a level which should be detectable
in the future with instruments like Herschel and ALMA.
\section{The Radio-FIR correlation\label{radio_fir}}
\begin{figure*}[t!]
\resizebox{\hsize}{!}{\includegraphics[clip=true]{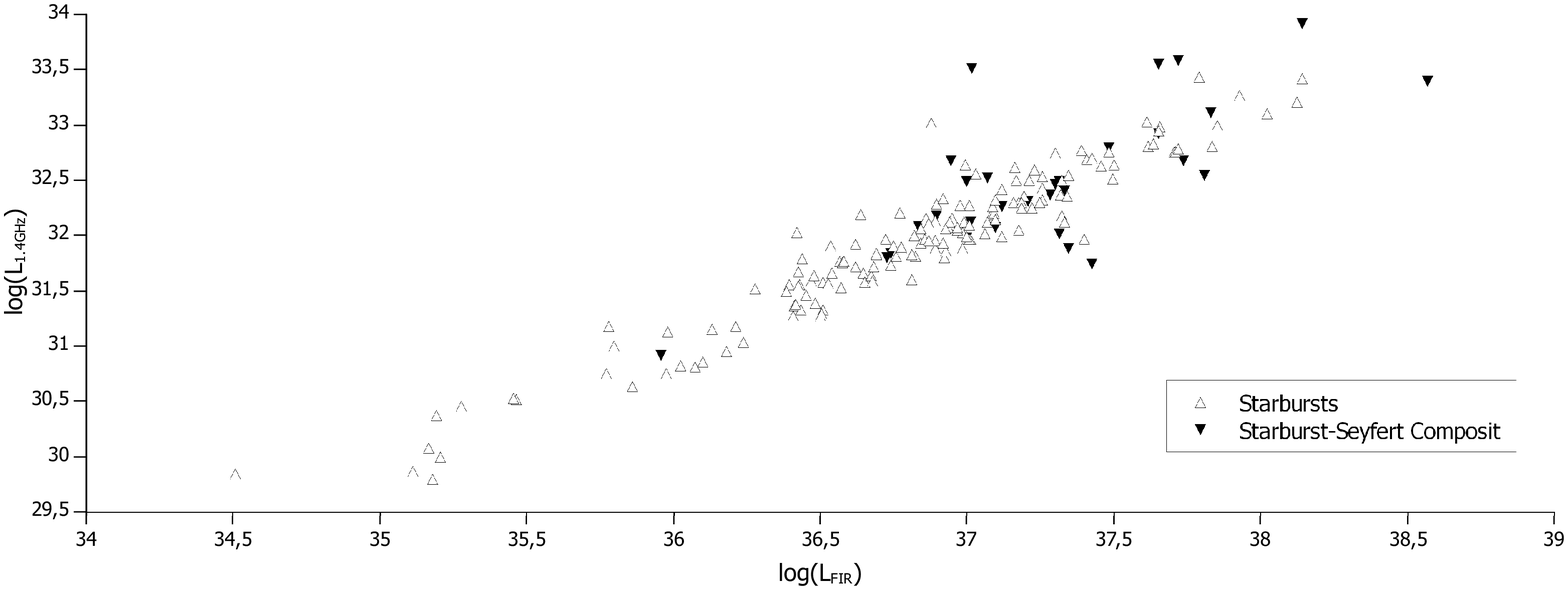}}
\caption{\footnotesize
FIR-radio correlation for a sample of local starburst galaxies
at $z\leq 0.03$, showing a linear correlation between the two
wavelengths. Luminosities are given in units of erg/s. The sample contains a significant fraction of sources
which, apart from the starburst behavior of the galaxy, reveal
Seyfert-like properties of the galaxy core. While those galaxies with
no Seyfert core identification are shown as open triangles, the
supposed Seyfert-starburst composites are shown as filled triangles.}
\label{fir_radio:fig}
\end{figure*}

The correlation between the total radio and far-infrared emission in
starburst galaxies is experimentally well established. As an example,
a sample of local starburst galaxies with redshifts below $z=0.03$ is
shown in Figure \ref{fir_radio:fig}. Far-infrared
emission arises when newly born, massive stars heat the surrounding
gas. Radio emission, on the other hand, is directly correlated to the
death of massive stars: electrons are
accelerated at supernova remnant shock fronts and lose their energy
via synchrotron radiation at radio wavelengths. Thus, from this very
general statement, a correlation between radio and far-infrared
emission is expected. A detailed modeling of the
correlation, however, remains difficult when trying to explain all observations. While a calorimetric
model, in which all electrons lose their entire energy to synchrotron
radiation seems feasible in the theoretical framework as presented by
\cite{voelk1989}, 
observations of
the radio spectral index in starburst galaxies show relatively flat
spectra: On average, radio spectra in the GHz range behave as $\sim
\nu^{-0.7}$. The synchrotron spectral index $\alpha_{\rm radio}$ and primary
particle spectral index $\alpha_p$ are directly connected as
\citep{rybicki_lightman1979} 
\begin{equation}
\alpha_p=2\cdot \alpha_{\rm radio}+1\,.
\end{equation}
Thus, a synchrotron spectral index of $\alpha_{\rm radio}\sim 0.7$
reveals a primary electron population with a spectral index of
$\alpha_{p}\sim 2.6$. In the calorimetric model, however, higher
energy electrons should lose all their energy, resulting in spectra as
steep as $E^{-3}$. An alternative model, based on a cosmic ray driven
wind model for starburst galaxies, is discussed
by \cite{becker_starbursts2009}. Here, the non-thermal radio flux,
induced by electron synchrotron losses,  turns out to be
only weakly dependent on the magnetic field of the starburst
($F_{\nu}\propto B^{0.12}$), even without the assumption of a
calorimeter. The question remains whether or not electrons can partly
escape the starburst galaxies and more detailed calculations have to
be performed in order to match all observational features.

An additional complication in the interpretation of electromagnetic
radiation from starburst galaxies is the apparent co-existence between
starburst galaxies and active Seyfert cores. Figure
\ref{fir_radio:fig} shows those galaxies that reveal an active core in
a starburst galaxy as
filled triangles. Thus, when the emission cannot be resolved
spatially, ambiguities can arise concerning the electromagnetic contribution from the
central activity and the starburst part of the galaxy.

\section{Hadronic interactions at $E>$~GeV \label{proton_proton:sec}}

A significant part of the total energy budget of a galaxy goes into
the acceleration of cosmic rays. It is known from the observation of
the charged cosmic ray flux at Earth, that the total energy carried by
Galactic cosmic rays with energies above $100$~GeV corresponds to a total luminosity of $\sim
10^{42}$~erg/s. The shock fronts of supernova remnants are one of the primary candidates
for the acceleration of cosmic rays up to $10^{15}$~eV or higher  \citep{biermann_cr1,biermann_pamela,biermann_wmap,biermann_composition}.
When interacting with
the local gas of the considered galaxy, a significant amount of the primary
cosmic rays' energy is put into the production of high-energy photons
and neutrinos. While neutrino telescopes have not yet reached the
sensitivity to detect diffuse emission from the Milky Way, gamma-ray
emission is detected at a level of $10^{39}$~erg/s. Thus, the
transport of cosmic rays in a galaxy is expected to play a significant
role considering the dynamics of the galaxy itself. The difficulty in
pin-pointing the sources of cosmic rays themselves lies in the
complexity of the transport of the charged particles through the
galactic magnetic field. For the Milky Way, the observed flux of
cosmic rays contains information about the total energy budget and the
spectral behavior after transport. However, no direct information on
the direction of the sources of cosmic rays or about the primary
spectra at injection can be deduced from observations. Concerning
starburst galaxies, compared to the high cosmic ray flux from the Milky
Way, any possible signal in charged cosmic rays would be
negligible. So, in general, in order to understand the role of cosmic
rays in a galaxy, the search for neutral secondaries like photons and
neutrinos from hadronic
interactions is one of the most interesting approaches.

High-energy photons and neutrinos are produced in hadronic interactions, in environments with a high flux
of high-energy cosmic rays $j_p(E_p)$ and a large target density
$n_H$ \citep{becker_starbursts2009}:
\begin{equation}
p\,p\rightarrow N(\pi^{+}/\pi^{-}/\pi^{0})+X\,.
\end{equation}
Here, $N(\pi^{+}/\pi^{-}/\pi^{0})$ denotes that multiple pions $N$ are created in the process.
Charged pions contribute to the production of high-energy
neutrinos, 
\begin{eqnarray}
\pi^{+}&\rightarrow& \mu^{+}\,\nu_{\mu}\rightarrow
e^{+}\,\nu_e\,\overline{\nu}_{\mu}\,\nu_\mu\\
\pi^{-}&\rightarrow& \mu^{-}\,\overline{\nu}_{\mu}\rightarrow e^{-}\,\overline{\nu}_e\,\nu_\mu\,\overline{\nu}_{\mu}\,.
\end{eqnarray}
Neutral pions produce high-energy photons, $\pi^{0}\rightarrow
\gamma\,\gamma$. 

Assuming that the dominant part of charged cosmic rays below
$10^{15}$~eV is accelerated
in supernova remnants, the target density directly at the acceleration
site is important as a local effect. At high hydrogen densities,
cosmic rays interact directly at the source and the neutral
interaction products reveal the spectral shape at injection. For
sources that have proton-proton optically thin environments, most
protons will escape the acceleration region and change their energy
spectrum due to transport effects. Then, interactions with the diffuse
gas in the galaxy serve as a tracer of the cosmic ray spectrum after
transport. Measurements concerning the transport of cosmic rays
through the Milky Way's magnetic field indicate that diffusion
steepens the injection spectrum by a factor of $\sim
E^{-0.3}-E^{-0.6}$ \citep{gupta_webber1989}. With the observed spectrum being close to
$E^{-2.7}$, the expected cosmic ray energy spectrum at injection is
expected to be close to $E^{-2.1}-E^{-2.4}$. At gamma-ray energies
above $\sim 300$~MeV, the photon spectrum roughly follows the charged
cosmic ray spectrum in the case of hadronic interactions. Thus, the
observation of the spectral behavior of gamma-rays from a galaxy can
help to identify if the interactions on average take place close to
the remnant, at injection, or rather in the interstellar medium, after transport.
While observations of our own
Galaxy indicate that propagation steepens the cosmic ray spectrum
before interactions take place, the two starburst galaxies with
observed gamma-ray spectra, M82 and NGC253, reveal a quite flat
spectral behavior of around $E^{-2.3}$. Thus, it is expected that a
large fraction of the cosmic ray interactions happen in the vicinity
of the sources.
\subsection{Gamma-ray observations of starburst galaxies}
As of today, there are six objects observed at gamma-ray energies,
where the emission appears to arise from hadronic interactions:
\begin{itemize}
\item {\bf M82 (Starburst galaxy)}\\
\item {\bf NGC253 (Starburst galaxy)}\\
\item {\bf The Milky Way}\\
\item {\bf Star-forming region 30 Doradus (Large Magellanic Cloud)}\\
\item {\bf NGC1068 (Starburst-Seyfert composite)}\\
\item {\bf NGC4945 (Starburst-Seyfert composite)}\\
\end{itemize}
Although the statistics with only six objects revealing hadronic
interactions is still low, it is worth to have a first look at what
kind of correlations to expect and to investigate if those show up in
this still very small sample. It is clear that there are many caveats when looking for correlations
 in this small sample: the only true starburst galaxies are M82 and NGC253. The star-forming region
30 Doradus is only a part of the dwarf galaxy. The Milky Way is a regular galaxy with contributions to the FIR emission from
stars that do not explode as supernovae. Thus, the Milky Way is not expected to lie on the FIR-radio correlation curve. The two starburst-Seyfert composites could have contributions from the active core in all wavelength ranges. Thus, future observations of a larger sample of pure starburst galaxies will have to confirm
or reject the results presented below.
\subsection{Interpretation of measured gamma-ray spectra as the
  product of hadronic interactions}
From Fermi- and IACT-observations, the gamma-ray flux from M82,
NGC253, the Milky Way and the star-forming region 30 Doradus in the LMC
are known \citep{fermi_starbursts2010}.
The measurements give a unique opportunity to draw conclusions about
the primary cosmic ray flux. 

First of all, the gamma-ray spectra for M82 and NGC253 appear to be
very flat, i.e.\ $E^{-2.3}$ and flatter. In the observed energy region
($>$~GeV), the hadronic gamma-ray flux follows the spectral behavior
of the primary cosmic rays. 

 Since the gamma-ray spectrum reproduces the same spectral shape as
 the interacting primary particles, most of the interactions must
 happen close to the acceleration region: while the injection spectrum
 is expected to be close to $E^{-2.0} -E^{-2.3}$  \citep{biermann_cr1,biermann_pamela,biermann_wmap,biermann_composition}, transport through the galactic magnetic field steepens the charged primaries' spectrum to $E^{-2.7}$. Interaction after transport into the ISM would therefore lead to a steep gamma-ray spectrum of close to  $E^{-2.7}$,

\begin{equation}
j_{\gamma} = \frac{dN_{\gamma}}{dE_{\gamma}~dt~dA_{Earth}}~,
\end{equation}
in units of GeV$^{-1}$~s$^{-1}$~cm$^{-2}$.

The number of pion-decay induced gamma-rays of a single SNR in a starburst per unit time, volume and energy is given \citep{kelner2006},
\begin{small}
\begin{eqnarray}
&\Phi_{\gamma}&(E_{\gamma})=n_H\cdot \nonumber\\
&\cdot& \int_{E_{\gamma}}^{\infty}{\sigma_{\rm{inel}}(E_p) \cdot j_p(E_p) \cdot F_{\gamma}\left(\frac{E_{\gamma}}{E_p},~E_p \right)~\frac{dE_p}{E_p}}\,.
\label{kelner_pp:eq}
\end{eqnarray}
\end{small}
Here, $n_H$ is the density of the ambient medium and
$\sigma_{\rm{inel}}(E_p)$ is the cross section of inelastic
proton-proton interactions\footnote{\cite{kelner2006} use the cosmic
  ray density, while here the cosmic ray flux is used.}. The function
$F_{\gamma}(x,~E_p)$ implies the number of photons in the energy
interval ($x, x~+~dx$) per collision and is a dimensionless
probability density distribution function. The flux of relativistic
protons at the source is given as 
\begin{small}
\begin{equation}
j_p(E_p)=a_p\cdot \Phi(E_p)\,.
\end{equation}
\end{small}
Here, $j_p$ is given per energy, time and area interval. The spectral
shape of the interacting proton flux is contained in the function
$\Phi(E_P)$ and can be expected to follow a power-law, $\Phi\sim
E^{-p}$, while $a_p$ is defined as the normalization factor, i.e.\ how
many protons there are per energy, time and area interval. This
normalization can be done at an arbitrary energy, we chose $E_p=1$~GeV to
normalize the spectra. Using other values yields the same
result when applying the same units consistently throughout the calculation.

To account for the total gamma-ray flux as observed at Earth, in units
of per time, area and energy interval,
Equation (\ref{kelner_pp:eq}) needs to be multiplied by the volume of the
interaction region $V$ of a single SNR and the number of SNRs $N_{\rm
  SNR}$, which scales directly with the SN rate $R_{\rm SN}$ of the
galaxy. Further, assuming isotropic emission, the
detected flux scales with $1/(4\,\pi\,d^2)$. The observed flux is then
directly connected to the produced density as
\begin{small}
\begin{eqnarray}
\left.\Phi\right|_{\oplus} &=& \Phi_{\gamma}(E_{\gamma}) \cdot V\cdot
N_{\rm SNR} \cdot \left(4\pi~d^2\right)^{-1}\\
&=& \frac{n_H \cdot V}{4\pi~d^2} \cdot \\
&\cdot&\int_{E_{\gamma}}^{\infty}\sigma_{\rm{inel}}(E_p) \cdot j_p(E_p) \cdot F_{\gamma}\left(\frac{E_{\gamma}}{E_p},~E_p \right)~\frac{dE_p}{E_p}\,.\nonumber
\end{eqnarray}
\end{small}
Assuming a proton flux at the source following a power-law with normalization  $a_p$ and a spectral shape $\Phi(E_p)$ yields
\begin{small}
\begin{eqnarray}
&\left.\Phi\right|_{\oplus}&  = a_{\gamma}\cdot\\
&\cdot&  \int_{E_{\gamma}}^{\infty} \sigma_{\rm{inel}}(E_p) \cdot \Phi_p(E_p) \cdot F_{\gamma}\left(\frac{E_{\gamma}}{E_p},~E_p \right)~\frac{dE_p}{E_p}\,,\nonumber
\end{eqnarray}
\end{small}
with
\begin{equation}
a_{\gamma} = \frac{a_p \cdot n_H \cdot V\cdot N_{\rm SNR}}{4\pi~d^2}\,.
\end{equation}
This factor needs to be fixed to a certain value to match the observations, and conclusions about the parameters on the right-hand side can be drawn. 

The proton spectral normalization, $a_p$ can be calculated using the
conservation of energy:
\begin{small}
\begin{eqnarray}
\int_{E_{min}}^{\infty}j_p\,E_p\,dE_p &=& a_p \cdot \int_{E_{min}}^{\infty} \Phi(E_p)\,E_p\,dE_p \nonumber\\
&=& \frac{W_p \cdot c}{V}\,.
\end{eqnarray}
\end{small}
where $W_p$ is the total proton energy budget of protons with a
minimum energy of $E_{\min}$, assuming the protons travel
approximately at the speed of light. Solving the equation for $a_p$ gives:
\begin{equation}
a_p = \frac{W_p \cdot c}{V\cdot \int_{E_{min}}^{\infty}\Phi(E_p)\, E_p\,.dE_p}\,,
\end{equation}
Thus, the normalization of the photon spectrum can be written as
\begin{equation}
a_{\gamma} = \frac{W_p \cdot c \cdot n_H\cdot N_{\rm SNR}}{4\pi d^2 \cdot\int_{E_{min}}^{\infty}{\Phi(E_p)~E_p~dE_p}}\,.
\end{equation}
Given the distance to the source and determining the average shape of
the primary particle spectrum from the shape of the gamma-ray spectrum, the product of the total
cosmic ray energy budget, the target density and the number of SNRs,
$W_p\cdot n_H\cdot N_{\rm SNR}$ determines the gamma-ray flux normalization. Alternatively, the cosmic ray energy density
$\rho_{\rm CR}=W_p/V$ can be used, and in this case, it is the product with
the total interacting mass, 
\begin{equation}
\rho_{\rm CR} \cdot M\cdot N_{\rm SNR}=\rho_{\rm CR}\cdot
n_H\cdot V\cdot N_{\rm SNR}
\end{equation}
is the determining factor. Thus, assuming
a constant cosmic ray energy density and a direct scaling of the number of
SNRs with the supernova rate $R_{\rm SN}$ would result in a direct
proportionality between the observed gamma-ray emission and the total
interacting gas mass. 

First Fermi results seem to suggest
that a correlation between the gamma-ray luminosity and the product of
the supernova rate and the total gas mass is present
\citep{fermi_starbursts2010}. 
However, the calculation has several
caveats: For once, it is not likely that the cosmic ray energy
density is constant for all starburst galaxies. In addition, it is
assumed here that all SNRs have the same cosmic ray energy spectrum,
although it is known that the spectrum actually strongly depends on
their age and the local environment of the sources, see
e.g.\ \cite{blasi2005} and references therein.
One might consider to rather search for a correlation between the
product of the SN rate and the gas density with the total gamma-ray
flux. However, even in this case, the above calculations rely on the
fact that the energy budget for each SNR is the same. However, it is
expected that it depends on the total mass of the stars and the energy
put into cosmic rays might not even be a constant for the same
progenitor star mass, but depend on the local environment. Further,
the spectral behavior of the cosmic rays might vary depending on the
local environment. 
And, finally, the average observed density of the
galaxy might not represent the average density in which the secondary
photons are produced.
\subsection{A correlation between radio and gamma-ray emission}
\begin{figure}[t!]
\resizebox{\hsize}{!}{\includegraphics[clip=true]{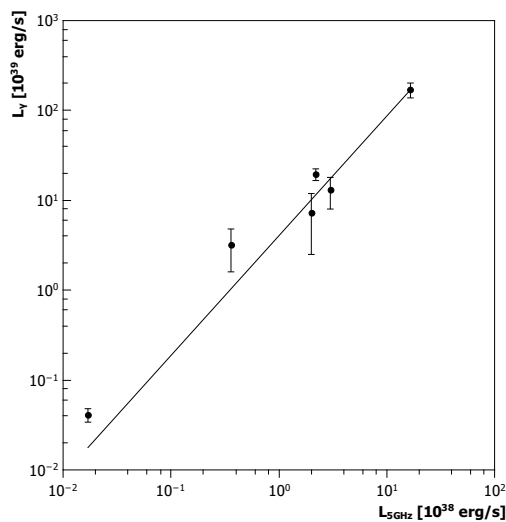}}
\caption{\footnotesize
Luminosities at gamma-ray energies as measured by Fermi versus radio
luminosities at $5$~GHz. Shown are the LMC, the Milky Way, M82,
NGC253, NGC1068 and NGC4945, using the data as summarized by
\cite{lenain2011}. The line represents a power-law fit to the
data. Due to the small number of sources, the error of the correlation
is still too large to make a quantitative statement. Qualitatively,
the result is compatible with a linear correlation.
}
\label{radio_gamma:fig}
\end{figure}

The non-thermal radio emission at GHz frequencies
is expected to come from synchrotron radiation of electrons
accelerated at supernova shock fronts. The gamma-ray emission, on the
other hand, is believed to have the same origin, following the
interpretation of the signal as the interaction of cosmic rays with the ambient medium. 
Thus, a correlation between the radio and gamma-ray
luminosities is expected to be present. Figure \ref{radio_gamma:fig}
shows the gamma-ray luminosity for the six objects versus their radio
luminosity. A clear trend of increased gamma-ray emission at enhanced
radio emission is observed. While the performed power-law fit is
compatible with a linear correlation of the two emission features, the
lack of statistics does not allow for the formulation of a
quantitative statement on this correlation. This might still be a
first hint that there is a connection between the two wavelengths,
which could further be used to study acceleration processes. 

\subsection{High-energy neutrinos}
Due to the co-production of high-energy photons and neutrinos,
hadronically produced gamma-rays are always accompanied by a neutrino
flux. Due to their low interaction probability, the detection of
neutrinos requires the use of kilometer-scale natural water or ice
reservoirs, see e.g.\ \cite{becker_review}. As of December 18, 2010, the first cubic-kilometer scale
neutrino detector IceCube was completed at the geographic South
Pole. The detection technique allows for the observation of the
entire northern hemisphere at a duty cycle of more than 99\%. The main
background are neutrinos produced in the Earth's atmosphere. Point
source searches can be performed for single sources as well as by
the stacking of a source list. If a contribution from unresolvable
sources is expected, which is the case for instance for several
classes of active galactic nuclei, a search for a diffuse flux can be
performed. Since astrophysical sources are expected to produce
relatively flat neutrino energy spectra for interactions at the source
($\sim E^{-2}-E^{-2.3}$), the explicit search for enhanced emission at
the highest energies reduces the background of the very steep
atmospheric neutrino flux $(\sim E^{-3.7})$. 

Concerning the case of starburst galaxies, the gamma-ray detection
from M82 gives a first concrete test case of the expected neutrino
flux from a point source. It turns out that the flux of M82 as a
single source is relatively low and it is not expected to be detected
within the first years of operation with IceCube. It may still be
possible to observe M82 after a longer period of measurement, as the
lifetime of IceCube is expected to be longer than 10 years. On the
other hand, the stacking of a source catalog of nearby starburst
galaxies strongly improves the detection significance. As the
significance roughly scales with the signal over the square root of
the background, adding more signal to the search helps to reveal the
signal over background. A first stacking search for starburst galaxies
has been performed with partially completed configurations of the
IceCube detector, see \cite{jens_turku,ic40_starbursts2010}. It is expected that within the
next few years, the limits can be improved using a detector of more
than twice the size. The detection of a diffuse flux from starburst
galaxies is rather challenging, since the maximum neutrino energy may
be as low as $10^{14}$~eV.


\section{Molecular ions as cosmic ray tracers \label{crtracers:sec}}
While protons with GeV energies and above contribute to the neutrino-
and photon flux of a galaxy, low-energy cosmic rays in the keV-GeV
range ionize the interstellar medium. In the Milky Way, the average
ionization level is observed to be of the order of $\zeta\approx
2\cdot 10^{-16}$~s$^{-1}$ \citep{gerin2010,neufeld2010}. Hydrogen
ionization immediately leads to the formation of $H_{2}^{+}$, which in
turn initiates the formation of larger molecules like $H_{3}^{+}$,
$OH^+$, $H_2O^+$ etc., see e.g.\ \citep{black1998} and references
therein. Those molecules can be observed by detectors like Herschel
and ALMA and can be used as direct tracers of cosmic ray ionization. 
The search for molecular ions at potential
cosmic ray acceleration sites with suitable targets might help to
trace the sources of cosmic rays and by that improve the understanding
of the role of cosmic rays in the dynamical processes of galaxies. 

In the Milky Way, several systems of supernova remnants and molecular
clouds (SNR-MC systems) have been detected at gamma-ray energies in the past years. In
particular, the detections of the sources W51C \citep{fermi_w51c}, W44
\citep{fermi_w44}, W28 \citep{fermi_w28}, IC443 \citep{fermi_ic443}
and W49B \citep{fermi_w49b}
with the Fermi Gamma-ray Space Telescope are best-fit with a hadronic
interaction model. The observed gamma-ray spectra can be used to
estimate the primary cosmic ray spectra at the source above GeV
energies. Extrapolating the spectrum down to below GeV energies then gives
the opportunity to perform calculations concerning interaction of
the low-energy part of the cosmic ray spectrum leading to the
ionization of the local medium. Details of the calculation are
presented in \cite{crtracers2011}. Due to the enhanced flux of cosmic
rays, the ionization level is expected to be enhanced by a few orders
of magnitude at the discussed SNR-MC systems. In such an environment,
the detection of line emission spectra from H$_{2}^{+}$ and H$_{3}^{+}$ would be the
most direct tracer for cosmic ray ionization. 

Line emission
from H$_{3}^{+}$ has been discussed previously in the context of cosmic ray ionization, and it represents one of the molecules observed in astrophysical contexts, see e.g.\ \citep{black1998} and references therein. With the launch of the
Herschel telescope, detailed observations of the abundance of H$_{3}^{+}$ is now possible in astrophysical environments. In \cite{indriolo_ic443}, for instance, an H$_{3}^{+}$ abundance corresponding to an ionization rate of $\sim 2\cdot 10^{-15}$/s was observed.

Although H$_{2}^{+}$ is the first product of cosmic ray ionization, it is
usually destroyed too quickly to produce significant line
emission. However, in an environment of extreme ionization level as it
seems to be the case for SNR-MC systems, the detection of H$_{2}^{+}$
seems to be possible and would provide a unique method to trace the
sources of cosmic rays.

Due to their high star formation rate, starburst galaxies are bound to
host a larger number of SNR-MC systems. In the future, it would
therefore be interesting to try to combine gamma-ray measurements with
the search for molecular line emission in order to be able to
pin-point the cosmic ray component of the galaxies.


\begin{acknowledgements}
The authors would like to thank S.\ Aalto, P.\ L.\ Biermann,
J.\ H.\ Black, S.\ Casanova, J.\ S. Gallagher, E.\ J\"utte, M.\ Olivo and
R.\ Schlickeiser for helpful and inspiring discussions.
\end{acknowledgements}
\bibliographystyle{aa}

\bigskip
\bigskip
\noindent {\bf DISCUSSION}

\bigskip
\noindent {\bf ARNON DAR:} Theoretical arguments rather point to a
correlation between the gamma-ray flux and the density of the galaxy,
and not to a scaling with the total gas mass.

\bigskip
\noindent {\bf JULIA BECKER:} Assuming a constant total energy budget
for all SNRs, this is correct, the scaling should be with the target
density of the interaction.
However, assuming a constant cosmic ray energy density gives
a scaling with the total gas mass. The difficulty with both arguments
is that there still are a lot of
assumptions in both statements: For instance, in both calculations,
all SNRs are assumed to have the same cosmic ray injection
spectrum. This is not the case in reality. In addition, the average
density of the galaxy does not necessarily represent the actual
density relevant for the interactions.

\bigskip
\noindent {\bf WOLFGANG KUNDT:} How do you avoid the production of
hard electrons from ionization?

\bigskip
\noindent {\bf JULIA BECKER:} The ionization cross sections decreases
rapidly with energy and can thus be neglected above GeV energies. This
is why below GeV energies, ionization processes dominate and above GeV
energies, hadronic interactions are prevailing.

\end{document}